\documentclass[twocolumn,english,pre,aps]{revtex4}
\usepackage[T1]{fontenc}
\usepackage[latin1]{inputenc}
\usepackage{amsmath}
\usepackage{graphicx}
\usepackage{amssymb}

\makeatletter
\def\vec#1{\mbox{\boldmath $\mathrm{#1}$}}
\def\Pm{\textrm{Pm}}
\def\Rm{\textrm{Rm}}
\def\RE{\textrm{Re}}
\def\Ha{\textrm{Ha}}

\usepackage{babel}
\makeatother

\begin{document}

\title{Pseudo--magnetorotational instability in a Taylor-Dean flow between
electrically connected cylinders}

\author{J\={a}nis Priede }

\email{j.priede@coventry.ac.uk}

\affiliation{Applied Mathematics Research Centre, Coventry University, Coventry
CV1 5FB, United Kingdom}

\begin{abstract}
We consider a Taylor-Dean-type flow of an electrically conducting
liquid in an annulus between two infinitely long perfectly conducting
cylinders subject to a generally helical magnetic field. The cylinders
are electrically connected through a remote, perfectly conducting
endcap, which allows a radial electric current to pass through the
liquid. The radial current interacting with the axial component of
magnetic field gives rise to the azimuthal electromagnetic force,
which destabilizes the base flow by making its angular momentum decrease
radially outwards. This instability, which we refer to as the pseudo--magnetorotational
instability (MRI), looks like an MRI although its mechanism is basically
centrifugal. In a helical magnetic field, the radial current interacting
with the azimuthal component of the field gives rise to an axial electromagnetic
force, which drives a longitudinal circulation. First, this circulation
advects the Taylor vortices generated by the centrifugal instability,
which results in a traveling wave as in the helical MRI (HMRI). However,
the direction of travel of this wave is opposite to that of the true
HMRI. Second, at sufficiently strong differential rotation, the longitudinal
flow becomes hydrodynamically unstable itself. For electrically connected
cylinders in a helical magnetic field, hydrodynamic instability is
possible at any sufficiently strong differential rotation. In this
case, there is no hydrodynamic stability limit defined in the terms
of the critical ratio of rotation rates of inner and outer cylinders
that would allow one to distinguish a hydrodynamic instability from
the HMRI. These effects can critically interfere with experimental
as well as numerical determination of MRI.
\end{abstract}

\pacs{47.20.Qr, 47.65.-d, 95.30.Lz}

\maketitle

\section{Introduction}

The magnetorotational instability (MRI) can account for the formation
of stars and entire galaxies in the accretion disks. For an object
to form, the matter circling around it has to slow down by transferring
its angular momentum outwards. The observed accretion rates suggest
the angular momentum transfer in the astrophysical disks to be turbulent
while the velocity distribution in them seems to be hydrodynamically
stable. A possible solution to this problem was suggested by Balbus
and Hawley \cite{Balbus-Hawley-1991,Balbus-Hawley-1998}, who pointed
out that a Keplerian velocity distribution in accretion disk can be
destabilized by a magnetic field in the process known as the MRI \cite{Velikhov-1959,Chandrasekhar-1960}.
This proposition has triggered a number of experimental studies trying
to reproduce MRI in laboratory \cite{Sisan-etal,Nature-2006}. The
main technical difficulty to such experiments is the magnetic Reynolds
number $\Rm$ that is required to be $\sim10$ at least. For a typical
liquid metal with the magnetic Prandtl number $\Pm\sim10^{-5}-10^{-6},$
this corresponds to a hydrodynamic Reynolds number $\RE=\Rm/\Pm\sim10^{6}-10^{7}$
\cite{Goodman-Ji-2002}. Thus, the base flow on which the MRI is
to be observed may be turbulent at such Reynolds numbers independently
of MRI as in the experiment of Sisan \emph{et al.} \cite{Sisan-etal}.
A possible solution to this problem was proposed by Hollerbach and
R\"udiger \cite{Hollerbach-Ruediger-2005}, who suggested that MRI
can take place in the Taylor-Couette (TC) flow at $\RE\sim10^{3}$
when the imposed magnetic field is helical rather than purely axial
as in the classical case. Theoretical prediction of this new type
of helical MRI (HMRI) was soon succeeded by a claim of its experimental
observation by Stefani \emph{et al.} \cite{Rued-apjl,Stefani-etal,Stefani-NJP}.
Subsequently, these experimental observations have been questioned
by Liu \emph{et al.} \cite{Liu-etal2006} who find no such instability
in their inviscid theoretical analysis of finite length cylinders
with insulating endcaps. In a more realistic numerical simulation,
Liu \emph{et al.} \cite{Liu-etal2007} confirm the experimental results,
though note that there is no MRI at the experimental parameters when
ideal TC boundary conditions are used. Szklarski \cite{Szklarski}
showed later that the ideal TC requires a slightly different parameters
for the HMRI to set in. Despite the numerical evidence, Liu \emph{et
al.} \cite{Liu-etal2007,Liu2008} suspected the observed phenomenon
to be a transient growth rather than a self-sustained instability.
This paper shows that the observation of a self-sustained instability
which looks like an MRI does not necessarily mean that the latter
is MRI.

Recently, we found that HMRI can be self-sustained and thus experimentally
observable in a system of sufficiently large axial extension because
there is not only convective but also absolute HMRI threshold \cite{Priede-Gerbeth2008}.
However, the comparison with the experimental results \cite{Rued-apjl,Stefani-etal,Stefani-NJP}
revealed that HMRI has been observed slightly beyond the range of
its absolute instability, where it is expected to be self-sustained
according to the ideal TC flow model. This discrepancy with the experimental
observations is probably due to the deviation of the real base flow
from the ideal TC flow used in the theoretical analysis. Such a deviation,
however, poses a major problem for the interpretation of experimental
results, especially for the identification of HMRI. Namely, the Rayleigh
line defining the hydrodynamic stability limit of the ideal TC flow
is used as a reference point to discriminate between a magnetically
modified Taylor vortex flow and HMRI. The latter two are hardly distinguishable
by the oscillation frequency, which varies weakly as the Rayleigh
line is crossed. The main problem is the hydrodynamic stability limit
of the real base flow, \textit{i.e.}, its actual Rayleigh line, which
may differ from that of the ideal TC flow. Therefore the latter cannot
be used for the interpretation of experimental results. This ambiguity
is not resolved by the direct numerical simulation of the problem
either even if there is a perfect agreement with the experiment. It
is because the notion of MRI is based on the ideal TC flow with a
fixed hydrodynamic stability limit, which is affected neither by the
end effects nor by the magnetic field. Unfortunately, this is the
case neither for experiments nor for numerical simulations. First,
there is an Ekman pumping at the endcaps, which can spread up to significant
but nevertheless limited distances into the base flow provided that
the latter is hydrodynamically stable. The Ekman circulation can be
reduced by using several independently rotating rings for the endcaps
\cite{Nature-2006} or by splitting the latter into two rings of
a certain size attached to the inner and outer cylinders, respectively
\cite{Szklarski}. Another important effect pointed out by Szklarski
and R\"udiger \cite{Szklarski-Ruediger2007}, which can significantly
affect the base flow, is related with the Hartmann layers forming
at the endcaps in axial magnetic field.

In this paper, we show that there may be additional effects in the
presence of a magnetic field when well conducting inner and outer
cylinders are electrically connected through an endcap as in the original
PROMISE experiment \cite{Rued-apjl,Stefani-etal,Stefani-NJP}. The
endcap acting in parallel with the Hartmann layer allows a radial
current to close through the liquid between the cylinders. The interaction
of radial current with axial magnetic field gives rise to an azimuthal
electromagnetic force, which reduces the velocity difference between
the endcap and the liquid above it. Depending on the strength of the
magnetic field, this electromagnetic force can render the profile
of azimuthal base flow centrifugally unstable. As a result, in axial
magnetic field, the instability can extend significantly beyond the
Rayleigh line similarly to the classical MRI. Moreover, in helical
magnetic field, the interaction of radial current with the azimuthal
component of magnetic field gives rise to an axial electromagnetic
force, which drives a longitudinal flow. First, this longitudinal
flow going upwards along the inner cylinder, where the azimuthal base
flow is centrifugally destabilized, advects Taylor vortices that results
in a traveling wave as in the HMRI. However, the direction in which
these Taylor vortices are advected is opposite to the direction of
travel of true HMRI wave. Second, for sufficiently large differential
rotation, longitudinal flow may become linearly unstable at any rotation
rate ratio.

The paper is organized as follows. In Sec. \ref{sec:prob-form} we
formulate the problem in the inductionless approximation. The base
flow for electrically connected cylinders is derived in Sec. \ref{sec:bstate}.
Section \ref{sec:pert} introduces the linear stability problem. Numerical
results for axial and helical magnetic fields are presented in Secs.
\ref{sub:Axial} and \ref{sub:Helical}, respectively. The paper is
concluded with a summary in Sec. \ref{sec:Summ}.

\section{\label{sec:prob-form}Problem formulation}

\begin{figure}
\begin{centering}
\includegraphics[height=0.7\columnwidth]{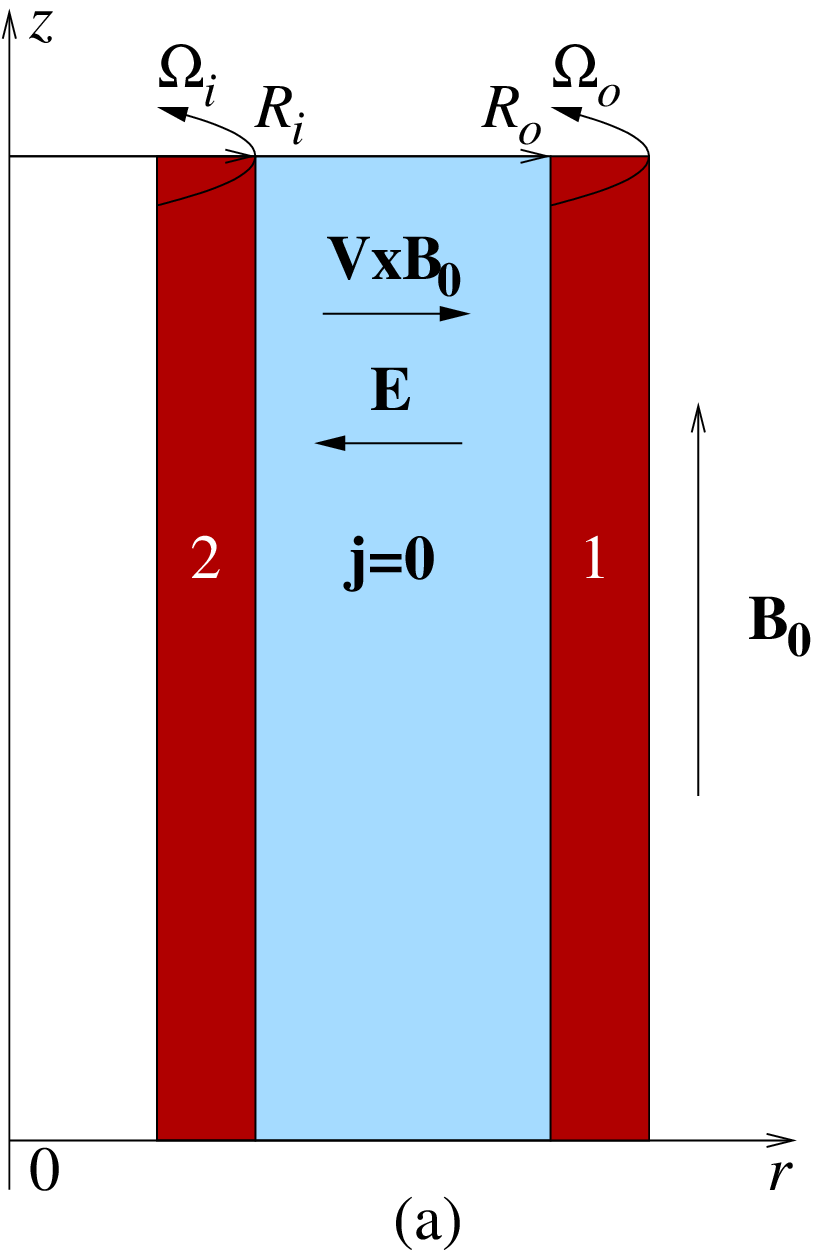} \includegraphics[height=0.7\columnwidth]{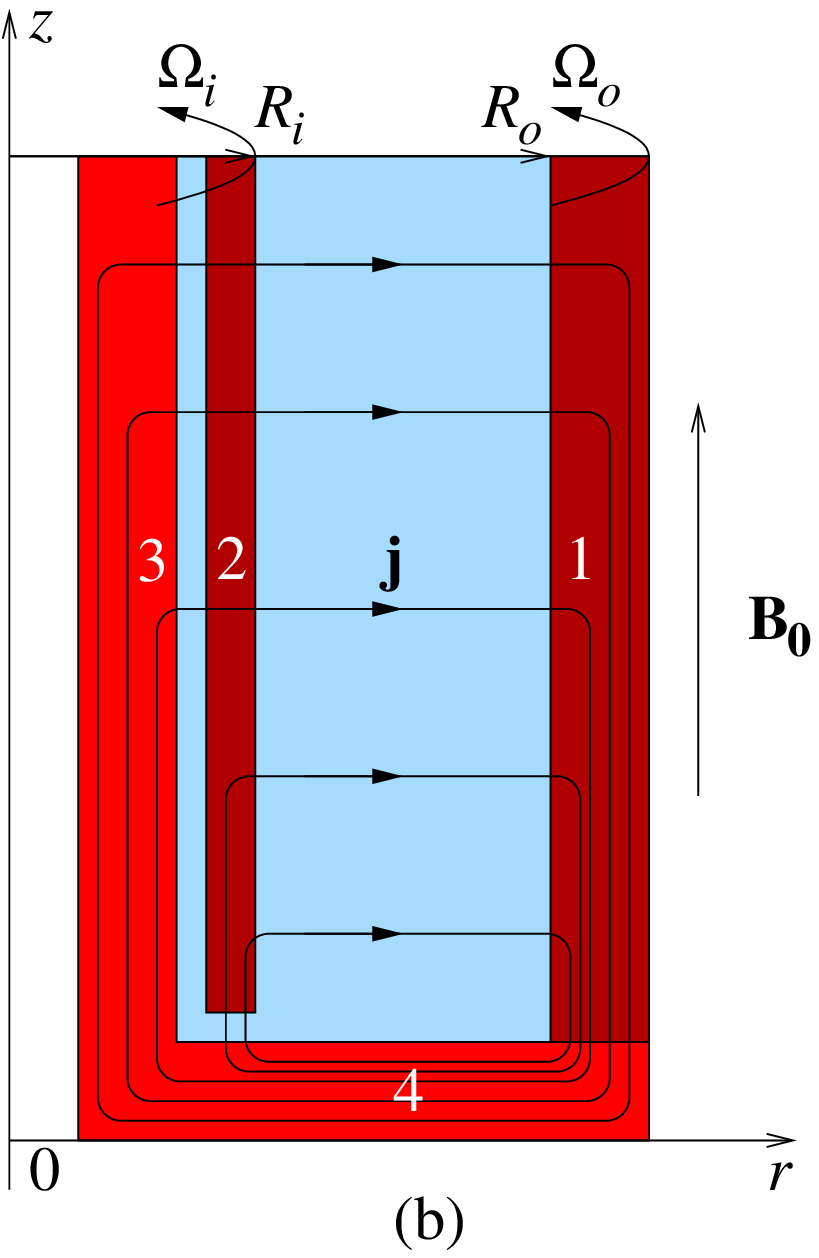}
\par\end{centering}

\caption{\label{fig:pseudo}(a) An ideal electrically uncoupled system and
(b) a real system with the inner and outer cylinders electrically
connected via the endcap and inner vessel wall; (1) outer cylinder,
(2) inner cylinder, (3) inner wall, (4) endcap. }

\end{figure}

Consider an incompressible fluid of kinematic viscosity $\nu$ and
electrical conductivity $\sigma$ filling the annulus between two
long, concentric cylinders with inner radius $R_{i}$ and outer radius
$R_{o}$ rotating with angular velocities $\Omega_{i}$ and $\Omega_{o}.$
The flow is subject to a generally helical, steady external magnetic
field $\vec{B}_{0}=B_{\phi}\vec{e}_{\phi}+B_{z}\vec{e}_{z}$ with
axial and azimuthal components $B_{z}=B_{0}$ and $B_{\phi}=\beta B_{0}R_{i}/r$
in cylindrical coordinates $(r,\phi,z),$ where $\beta$ is a dimensionless
parameter characterizing the geometrical helicity of the field. The
fluid is supposed to be poorly conducting so that the induced magnetic
field is negligible with respect to the imposed one. This corresponds
to the so-called inductionless approximation, which holds for the
HMRI characterized by small magnetic Reynolds number $\Rm=\mu_{0}\sigma v_{0}L\ll1,$
where $\mu_{0}$ is the permeability of vacuum, $v_{0}$ and $L$
are the characteristic velocity and length scale \cite{Priede-etal-2007}.
The velocity of fluid flow $\vec{v}$ is governed by the Navier-Stokes
equation with electromagnetic body force \begin{equation}
\frac{\partial\vec{v}}{\partial t}+(\vec{v}\cdot\vec{\nabla})\vec{v}=-\frac{1}{\rho}\vec{\nabla}p+\nu\vec{\nabla}^{2}\vec{v}+\frac{1}{\rho}\vec{j}\times\vec{B}_{0},\label{eq:N-S}\end{equation}
 where the induced current follows from Ohm's law for a moving medium
\begin{equation}
\vec{j}=\sigma\left(\vec{E}+\vec{v}\times\vec{B}_{0}\right).\label{eq:Ohm}\end{equation}
In addition, we assume that the characteristic time of velocity variation
is much longer than the magnetic diffusion time, $\tau_{0}\gg\tau_{m}=\mu_{0}\sigma L^{2},$
that leads to the quasi-stationary approximation, according to which
$\vec{\nabla}\times\vec{E}=0$ and $\vec{E}=-\vec{\nabla}\Phi,$ where
$\Phi$ is the electrostatic potential. Mass and charge conservation
require $\vec{\nabla}\cdot\vec{v}=\vec{\nabla}\cdot\vec{j}=0.$

\section{\label{sec:bstate}Base state}

An ideal, axially unbounded system shown in Fig. \ref{fig:pseudo}(a)
admits a translationally invariant base state with purely azimuthal
velocity distribution $\vec{v}_{0}(r)=\vec{e}_{\phi}v_{0}(r).$ Such
a flow in axial magnetic field induces a radial electric field, which
gives rise only to the potential difference between the inner and
outer cylinder but no radial current is induced because of the charge
conservation. Thus, in an ideal system, the magnetic field affects
the stability of the base flow without altering the latter, which
is the main premise underlying MRI. In reality both cylinders may
not be completely electrically decoupled from each other. For example,
in an axially bounded system such a coupling may be provided by an
electrically conducting endcap serving as a closing circuit between
the inner and outer cylinders. This corresponds to the principal setup
of the PROMISE experiment illustrated in Fig. \ref{fig:pseudo}(b).
The liquid metal in the narrow gap separating the inner cylinder $(2)$
from the endcap $(4)$ and the inner wall $(3)$, which both form
a solid well-conducting vessel together with the outer cylinder, provides
a sliding contact. This allows a radial electric current to pass through
the liquid between the outer and inner cylinder and then to close
either directly through the endcap or via the inner wall as sketched
in Fig. \ref{fig:pseudo}(b). Note that this setup is analogous to
the homopolar generator also known as the Faraday disk.

In the following, an axially uniform radial current $\vec{j}_{0}=\vec{e}_{r}j_{0}(r)$
is supposed to pass through the liquid and close through a remote
endcap. Note that this current is generated by the differential rotation
of cylinders in axial magnetic field rather than applied externally
as it is planned in the so-called Kurchatov MRI experiment \cite{Velikhov-2006,Khalzov-2006}.
Our main assumption is that the system is sufficiently extended so
that an axially uniform base state can develop sufficiently far away
from the ends as in the classical TC setup. Thus, we neglect any direct
effect of the endcap on the base flow, which is affected only by an
axially uniform radial current passing through the liquid. The charge
conservation yields $j_{0}(r)=J_{0}/r,$ where $J_{0}$ is a constant
that will be determined later by specifying the connection between
the cylinders. First, the interaction of radial current with axial
magnetic field gives rise to the azimuthal electromagnetic force,
which affects the profile of azimuthal velocity. The latter is governed
by the $\phi$ component of Eq. (\ref{eq:N-S})\[
\frac{1}{r^{2}}\frac{d}{dr}\left[r^{3}\frac{d}{dr}\left(\frac{v_{0}}{r}\right)\right]=\frac{J_{0}B_{0}}{\nu\rho}\frac{1}{r},\]
whose solution can be written as $v_{0}(r)=\bar{v}_{0}(r)-J_{0}B_{0}/(2\nu\rho)\tilde{v}_{0}(r),$
where $\bar{v}_{0}(r)$ and $\tilde{v}_{0}(r)$ are the profiles of
the classical Couette and electromagnetically driven Dean \cite{Chandrasekhar-1961}
flows, \[
\bar{v}_{0}(r)=r\frac{R_{o}^{2}\Omega_{o}-R_{i}^{2}\Omega_{i}}{R_{o}^{2}-R_{i}^{2}}+\frac{1}{r}\frac{\Omega_{o}-\Omega_{i}}{R_{o}^{-2}-R_{i}^{-2}},\]
\[
\tilde{v}_{0}(r)=r\frac{R_{o}^{2}\ln R_{o}-R_{i}^{2}\ln R_{i}}{R_{o}^{2}-R_{i}^{2}}+\frac{1}{r}\frac{\ln R_{o}-\ln R_{i}}{R_{o}^{-2}-R_{i}^{-2}}-r\ln r.\]

Linear stability of such a Taylor-Dean (TD) flow in purely axial magnetic
field has been considered by Szklarski and R\"udiger \cite{Szklarski-Ruediger2007}.
In contrast to us, they regard the Dean component of the flow to be
independent of the Couette one, but ignore that the current driving
the former is induced by the latter, i.e., the differential rotation
of the cylinders.

Second, in a helical magnetic field, radial current interacting also
with the azimuthal component of magnetic drives a longitudinal flow
$w_{0}(z)$ governed by the $z$ component of Eq. (\ref{eq:N-S}),
\[
\frac{1}{r}\frac{d}{dr}\left(r\frac{dw_{0}}{dr}\right)=\frac{1}{\nu\rho}\left(\frac{\partial p_{0}}{\partial z}-\frac{J_{0}B_{0}\beta_{i}}{r^{2}}\right).\]
The solution can be presented as \[
w_{0}(r)=-\frac{P_{0}}{4\rho\nu}\tilde{w}_{0,1}(r)-\frac{J_{0}B_{0}\beta_{i}}{2\rho\nu}\tilde{w}_{0,2}(r),\]
where $\beta_{i}=\beta R_{i}$ and $\tilde{w}_{0,1}(r)$ and $\tilde{w}_{0,2}(r)$
are the parts of flow driven by the pressure gradient and by electromagnetic
force\begin{multline*}
\tilde{w}_{0,1}(r)=\frac{R_{o}^{2}/\ln R_{o}-R_{i}^{2}/\ln R_{i}}{1/\ln R_{o}-1/\ln R_{i}}+\frac{R_{o}^{2}-R_{i}^{2}}{\ln R_{o}-\ln R_{i}}\ln r-r^{2},\\
\tilde{w}_{0,2}(r)=\ln R_{o}\ln R_{i}-\ln\left(R_{o}R_{i}\right)\ln r+\ln^{2}r.\end{multline*}
 The axial pressure gradient $P_{0}=\partial p_{0}/\partial z,$ which
is constant for a longitudinally uniform flow, is related to the electromagnetically-driven
part of the flow by the flow rate conservation $\int_{R_{i}}^{R_{o}}w_{0}(r)rdr=0$
yielding $P_{0}=-2J_{0}B_{0}\beta_{i}K_{0},$ where\[
K_{0}=\int_{R_{i}}^{R_{o}}\tilde{w}_{0,2}(r)rdr/\int_{R_{i}}^{R_{o}}\tilde{w}_{0,1}(r)rdr\]
results in a simple but long analytic expression which is skipped
here. Eventually, we obtain $w_{0}(r)=J_{0}B_{0}\beta_{i}/(2\rho\nu)\tilde{w}_{0}(r),$
where $\tilde{w}_{0}(r)=K_{0}\tilde{w}_{0,1}(r)-\tilde{w}_{0,2}(r)$
depends on the geometry only. In order to determine the last unknown
quantity $J_{0},$ we need to specify how the inner and outer cylinders
are connected to each other by the endcap. In the following, we focus
on the experimental configuration shown in Fig. \ref{fig:pseudo}(b),
where the outer cylinder $(1)$ forms a solid body together with the
endcap $(4)$ and inner wall $(3),$ while the inner cylinder $(2)$
is separated from the endcap and the inner wall by a relatively thin
gap filled with the liquid metal, which serves as a sliding contact.
First, we integrate Ohm's law {[}Eq. (\ref{eq:Ohm})] over the liquid
gap giving us the radial voltage drop between the inner and outer
cylinders\begin{equation}
\Phi_{o}-\Phi_{i}=B_{o}\int_{R_{i}}^{R_{o}}\bar{v}_{0}(r)dr-J_{0}\left(\frac{1}{\sigma}\ln\frac{R_{0}}{R_{i}}+\frac{B_{0}^{2}}{2\nu\rho}\mathcal{I}\right),\label{eq:dphi0}\end{equation}
where $\mathcal{I}=\int_{R_{i}}^{R_{o}}\left[\tilde{v}_{0}(r)+\beta_{i}^{2}\tilde{w}_{0}(r)/r\right]dr$
represents another long analytic expression. Second, since there is
no axial voltage drop along perfectly conducting cylinders, the same
radial voltage drop can be found alternatively by integrating Ohm's
law radially from $R_{i}$ to $R_{o}$ over the endcap, which is also
assumed to be perfectly conducting,\begin{equation}
\Phi_{o}-\Phi_{i}=\frac{1}{2}B_{o}\Omega_{o}\left(R_{o}^{2}-R_{i}^{2}\right)+J_{0}\mathcal{R},\label{eq:dphi1}\end{equation}
where $\mathcal{R}$ is a phenomenological parameter introduced to
account for effective linear resistance of the sliding liquid-metal
contact between the inner cylinder and the endcap. Substituting this
into Eq. (\ref{eq:dphi0}) we obtain\begin{eqnarray}
J_{0} & = & B_{0}\left(\Omega_{o}-\Omega_{i}\right)\left(\frac{\ln R_{o}-\ln R_{i}}{R_{o}^{-2}-R_{i}^{-2}}+\frac{R_{i}^{2}}{2}\right)\nonumber \\
 &  & \times\left(\mathcal{R}+\frac{1}{\sigma}\ln\frac{R_{o}}{R_{i}}+\frac{B_{0}^{2}}{2\nu\rho}\mathcal{I}\right)^{-1},\label{eq:J0}\end{eqnarray}
 which is the last quantity defining the base state.

Now it remains to estimate the resistance $\mathcal{R}$ introduced
in Eq. (\ref{eq:dphi1}) for the setup shown in Fig. \ref{fig:pseudo}(b)
that is described in detail in Refs. \cite{Stefani-etal} and \cite{Stefani-NJP}.
As seen, there are two parallel paths for the current to connect between
the inner cylinder $(2)$ and the endcap $(4)$. First, the current
can connect directly over the vertical gap of $\approx10$~mm width
between the inner cylinder and the endcap. Second, the current can
connect over the annular gap of $4$~mm width between the inner cylinder
$(2)$ and the inner wall $(3)$ and then pass along the latter towards
the endcap $(4)$. Because of a much larger contact area, the effective
resistance of the second path is obviously much smaller than that
of the first one, which thus may be neglected in this parallel connection.
On the other hand, the gap width of the second path is by an order
of magnitude smaller than the $40$~mm width of whole liquid layer
between the inner and outer cylinders. Thus, the resistance of the
second path may be neglected with respect to that of the whole liquid
layer, which is connected in series with the latter. In the following,
we assume $\mathcal{R}=0$ that supposes a negligible contact resistance
between the inner and outer cylinders, which appears to be a good
approximation to this PROMISE setup. The limit of $\mathcal{R}\rightarrow\infty$
corresponds to the classical case of electrically decoupled cylinders.
Note that in Eq. (\ref{eq:J0}) $\mathcal{R}$ stands next to the
electromagnetic term $\sim B_{0}^{2}$ implying that even a finite
$\mathcal{R}$ may become negligible in sufficiently strong magnetic
field. In addition, note that the actual PROMISE setup is considerably
more complex than this simple model. In particular, we assume that
the sidewalls are perfectly conducting with respect to the liquid
metal, whereas the conductivity of Copper sidewalls is only 13 times
higher than that of the GaInSn eutectic alloy used in the experiment.
Although our model is relatively rough, it can still highlight some
principal effects overlooked by more elaborate numerical models.

\begin{figure*}
\begin{centering}
\includegraphics[width=0.45\textwidth]{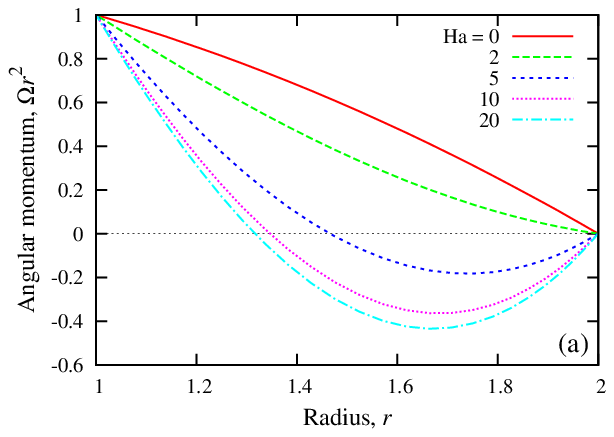} \includegraphics[width=0.45\textwidth]{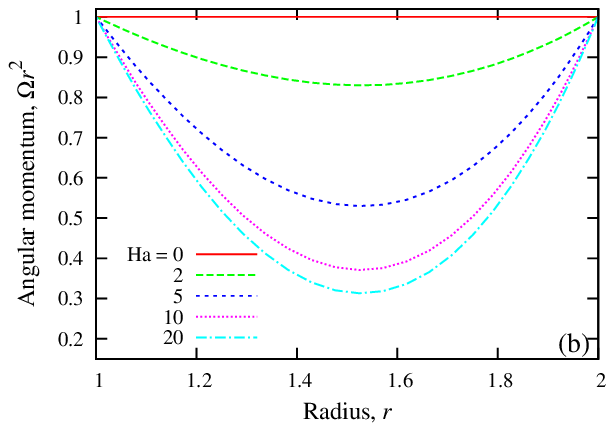}\\
\includegraphics[width=0.45\textwidth]{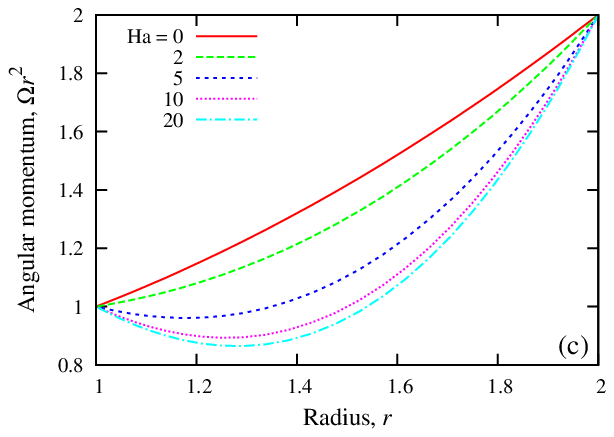}
\par\end{centering}

\caption{\label{fig:mphi-bt0}Angular momentum profiles for (a) $\mu=0,$ (b)
$\mu=0.25,$ (c) $\mu=0.5$ and various Hartmann numbers for cylinders
with $\lambda=2$ in axial magnetic field $(\beta=0)$ with no contact
resistance $(\bar{\mathcal{R}}=0).$ }

\end{figure*}

\section{\label{sec:pert}Perturbed state}

We consider a perturbed state \[
\left\{ \begin{array}{c}
\vec{v},p\\
\vec{j},\Phi\end{array}\right\} (\vec{r},t)=\left\{ \begin{array}{c}
\vec{v}_{0},p_{0}\\
\vec{j}_{0},\Phi_{0}\end{array}\right\} (r)+\left\{ \begin{array}{c}
\vec{v}_{1},p_{1}\\
\vec{j}_{1},\Phi_{1}\end{array}\right\} (\vec{r},t)\]
where $\vec{v}_{1},$ $p_{1},$ $\vec{j}_{1},$ and $\Phi_{1}$ present
small-amplitude perturbations for which Eqs. (\ref{eq:N-S}) and (\ref{eq:Ohm})
after linearization take the form \begin{eqnarray}
\frac{\partial\vec{v}_{1}}{\partial t} & + & (\vec{v}_{1}\cdot\vec{\nabla})\vec{v}_{0}+(\vec{v}_{0}\cdot\vec{\nabla})\vec{v}_{1}\nonumber \\
 & = & -\frac{1}{\rho}\vec{\nabla}p_{1}+\nu\vec{\nabla}^{2}\vec{v}_{1}+\frac{1}{\rho}\vec{j}_{1}\times\vec{B}_{0},\label{eq:v1}\\
\vec{j}_{1} & = & \sigma\left(-\vec{\nabla}\Phi_{1}+\vec{v}_{1}\times\vec{B}_{0}\right).\label{eq:j1}\end{eqnarray}
In this paper, we focus on axisymmetric perturbations, which are typically
more unstable than nonaxisymmetric ones for TC flow \cite{Rued-ANN},
however this is not always the case for the conventional TD flow \cite{Chen-1993}.
Analysis of nonaxisymmetric perturbations for an electromagnetically
driven TD flow is outside the scope of this paper. In the axisymmetric
case, the solenoidity constraints are satisfied by meridional stream
functions for fluid flow and electric current as \[
\vec{v}=v\vec{e}_{\phi}+\vec{\nabla}\times(\psi\vec{e}_{\phi}),\qquad\vec{j}=j\vec{e}_{\phi}+\vec{\nabla}\times(h\vec{e}_{\phi}).\]
Note that $h$ is the azimuthal component of the induced magnetic
field, which is used subsequently instead of $\Phi$ for the description
of the induced current. Thus, we effectively retain the azimuthal
component of the induction equation to describe meridional components
of the induced current, while the azimuthal current is related explicitly
to the radial velocity. In addition, for numerical purposes, we introduce
also the vorticity $\vec{\omega}=\vec{\nabla}\times\vec{v}=\omega\vec{e}_{\phi}+\vec{\nabla}\times(v\vec{e}_{\phi})$
as an auxiliary variable. The perturbation is sought in the normal-mode
form \[
\left\{ v_{1},\omega_{1,}\psi_{1},h_{1}\right\} (\vec{r},t)=\left\{ \hat{v},\hat{\omega},\hat{\psi},\hat{h}\right\} (r)e^{\gamma t+ikz},\]
where $\gamma$ is, in general, a complex growth rate and $k$ is
the axial wave number. Henceforth, we proceed to dimensionless variables
by using $R_{i},$ $R_{i}^{2}/\nu,$ $R_{i}\Omega_{i},$ $B_{0},$
and $\sigma B_{0}R_{i}\Omega_{i}$ as the length, time, velocity,
magnetic field, and current scales, respectively. The nondimensionalized
governing equations are\begin{eqnarray}
\gamma\hat{v} & = & D_{k}\hat{v}-\RE ik\left(w_{0}\hat{v}-r^{-1}\left(rv_{0}\right)'\hat{\psi}\right)+\Ha^{2}ik\hat{h},\label{eq:vhat}\\
\gamma\hat{\omega} & = & D_{k}\hat{\omega}-\RE ik\left(w_{0}\hat{\omega}+r\left(r^{-1}w_{0}'\right)'\hat{\psi}-2r^{-1}v_{0}\hat{v}\right)\nonumber \\
 &  & -\Ha^{2}ik\left(ik\hat{\psi}+2\beta r^{-2}\hat{h}\right),\label{eq:omghat}\\
0 & = & D_{k}\hat{\psi}+\hat{\omega},\label{eq:psihat}\\
0 & = & D_{k}\hat{h}+ik\left(\hat{v}-2\beta r^{-2}\hat{\psi}\right),\label{eq:hhat}\end{eqnarray}
where $D_{k}f\equiv r^{-1}\left(rf'\right)'-(r^{-2}+k^{2})f$ and
the prime stands for $d/dr;$ $\RE=R_{i}^{2}\Omega_{i}/\nu$ and $\Ha=R_{i}B_{0}\sqrt{\sigma/(\rho\nu)}$
are the Reynolds and Hartmann numbers, respectively. Boundary conditions
for the flow perturbation and the electric stream function on the
perfectly conducting inner and outer cylinders at $r=1$ and $r=\lambda,$
respectively, are $\hat{v}=\hat{\psi}=\hat{\psi}'=(r\hat{h})'=0.$

The governing Eqs. (\ref{eq:vhat})--(\ref{eq:hhat}) for perturbation
amplitudes were solved in the same way as in Refs. \cite{Priede-etal-2007}
and \cite{Priede-Gerbeth2008} by using a spectral collocation method
on a Chebyshev-Lobatto grid with a typical number of internal points
$N=32,$ which ensured the accuracy of about five digits.

The dimensionless azimuthal and axial velocity components of the base
flow\begin{eqnarray}
v_{0}(r) & = & \bar{v}_{0}(r)+\frac{1}{2}\Ha^{2}\bar{J}_{0}\tilde{v}_{0}(r),\label{eq:v0n}\\
w_{0}(r) & = & \frac{1}{2}\Ha^{2}\bar{J}_{0}\beta\tilde{w}_{0}(r),\label{eq:w0n}\end{eqnarray}
follow straightforwardly from the corresponding dimensional counterparts
when $R_{i}$ and $\Omega_{i}$ are replaced by $1,$ $R_{o}$ and
$\Omega_{o}$ by $\lambda=R_{o}/R_{i}$ and $\mu=\Omega_{o}/\Omega_{i},$
respectively, and $\beta_{i}$ by $\beta;$ the dimensionless counterpart
of $J_{0}$ is \begin{eqnarray}
\bar{J}_{0} & = & \frac{J_{0}}{\sigma B_{0}\Omega_{i}R_{i}^{2}}=\left(\mu-1\right)\left(\frac{\ln\lambda}{\lambda^{-2}-1}+\frac{1}{2}\right)\nonumber \\
 &  & \times\left(\bar{\mathcal{R}}+\ln\lambda+\frac{1}{2}\Ha^{2}\bar{\mathcal{I}}\right)^{-1},\label{eq:J0n}\end{eqnarray}
where $\bar{\mathcal{R}}=\mathcal{R}\sigma$ and $\bar{\mathcal{I}}=\mathcal{I}/R_{i}^{2}$
are the dimensionless counterparts of $\mathcal{R}$ and $\mathcal{I},$
respectively. As seen from Eqs. (\ref{eq:v0n})-- (\ref{eq:J0n}),
for $\Ha\gg1$ velocity profiles tend to asymptotic solutions which,
as noted above, depend neither on the contact resistance $\bar{\mathcal{R}}$
nor on the magnetic field strength \begin{eqnarray}
v_{0}(r) & \approx & \bar{v}_{0}(r)+\tilde{J}_{0}\tilde{v}_{0}(r),\label{eq:v0a}\\
w_{0}(r) & \approx & \tilde{J}_{0}\beta\tilde{w}_{0}(r),\label{eq:w0a}\end{eqnarray}
where $\tilde{J}_{0}=\left(\mu-1\right)\left(\frac{\ln\lambda}{\lambda^{-2}-1}+\frac{1}{2}\right)/\bar{\mathcal{I}}.$

\section{Numerical results}

\subsection{\label{sub:Axial}Axial magnetic field}

\begin{figure*}
\begin{centering}
\includegraphics[width=0.45\textwidth]{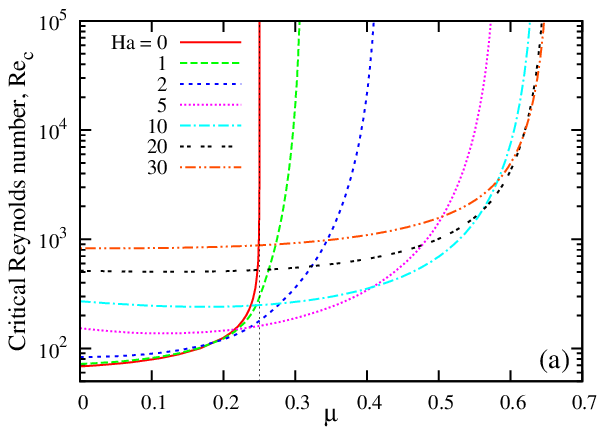} \includegraphics[width=0.45\textwidth]{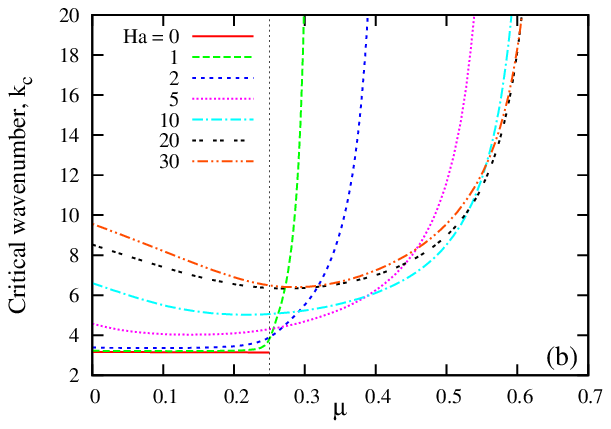}
\par\end{centering}

\caption{\label{fig:Rec-mu-bt0}(a) Critical Reynolds number $Re_{c}$ and
(b) wave number $k_{c}$ versus $\mu$ at various Hartmann numbers
for cylinders with $\lambda=2$ and no contact resistance $(\bar{\mathcal{R}}=0)$
in purely axial magnetic field.}

\end{figure*}

We start with an axial magnetic field $(\beta=0),$ for which the
base flow is purely azimuthal. The profiles of angular momentum $rv_{0}(r)$
are shown in Fig. \ref{fig:mphi-bt0} for several cylinder rotation
rate ratios $\mu$ and various Hartmann numbers. For $\mu=0,$ shown
in Fig. \ref{fig:mphi-bt0}(a), which corresponds only to the inner
cylinder rotating, the profile without the magnetic field is centrifugally
unstable with the angular momentum decreasing radially outward. In
this case, the magnetic field slows down the overall rotation rate
of the liquid making the angular momentum decrease faster at the inner
cylinder that may result even in the reversal of the sense of liquid
rotation at the outer cylinder when the magnetic field is sufficiently
strong. This effect is due to the magnetic field trying to eliminate
the differential rotation between the liquid and the endcap, which
is attached to the outer cylinder and thus rotates with a lower angular
velocity than the liquid above it as long as $\mu<1.$ A similar effect
can also be observed in Fig. \ref{fig:mphi-bt0}(b) for $\mu=0.25,$
which without the magnetic field corresponds to a marginally stable
state with a constant angular momentum distribution. In this case,
the magnetic field again retards the liquid rotation so rending the
distribution of angular momentum centrifugally unstable at the inner
cylinder and stable at the outer one. For $\mu=0.5$ shown in Fig.
\ref{fig:mphi-bt0}(c), the profile without magnetic field is centrifugally
stable with the angular momentum increasing radially outward. However,
a strong enough magnetic field changes the distribution of the angular
momentum at the inner cylinder from radially increasing to decreasing
one so rending the profile centrifugally unstable.

This is confirmed by the critical Reynolds number plotted against
$\mu$ for various Hartmann numbers in Fig. \ref{fig:Rec-mu-bt0}(a)
with the corresponding critical wave numbers shown in Fig. \ref{fig:Rec-mu-bt0}(b).
As seen in Fig. \ref{fig:Rec-mu-bt0}(a), without magnetic field $(\Ha=0),$
the critical Reynolds number tends to infinity as $\mu$ approaches
the Rayleigh line $\mu_{c}=\lambda^{-2}=0.25$ defined by $d\left(r^{2}\Omega\right)/dr=0,$
at which the profile of angular momentum becomes centrifugally stable.
As the magnetic field is increased, the instability starts to extend
beyond the Rayleigh line reaching $\mu\approx0.65$ at sufficiently
large Hartmann numbers. Although this extension of the instability
beyond the Rayleigh line may look like an MRI, it has a principally
different physical mechanism. Namely, in the MRI, the magnetic field
destabilizes the flow without altering it, whereas here the magnetic
field does alter the base flow by rendering it centrifugally unstable
as discussed above. Moreover, the standard MRI in axial magnetic field
is not captured by the inductionless approximation $(\Pm=0)$ used
here \cite{Herron-Goodman-2006}. Thus, in axial magnetic field,
this centrifugal instability occurring beyond the Rayleigh line can
easily be distinguished from the true MRI.

\subsection{\label{sub:Helical}Helical magnetic field}

\begin{figure*}
\begin{centering}
\includegraphics[width=0.45\textwidth]{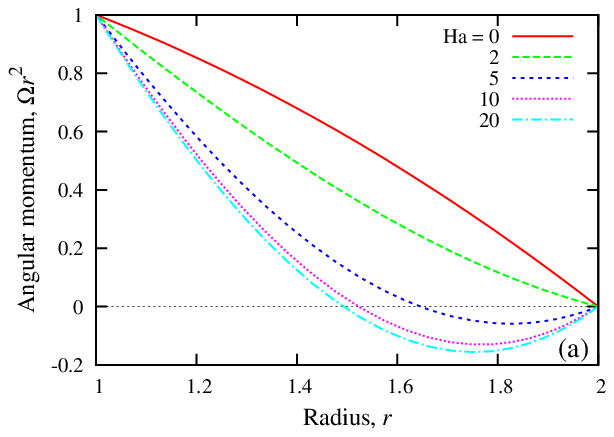} \includegraphics[width=0.45\textwidth]{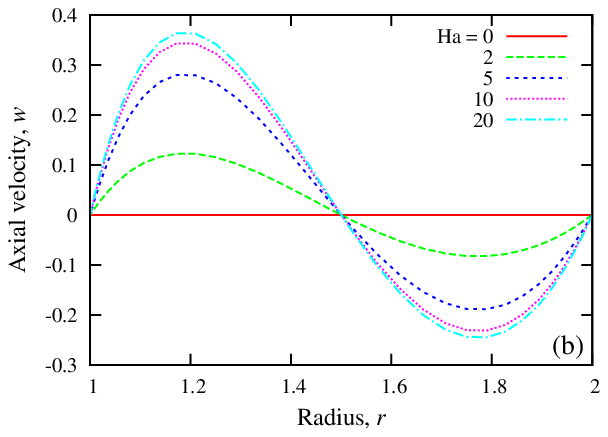}\\
\includegraphics[width=0.45\textwidth]{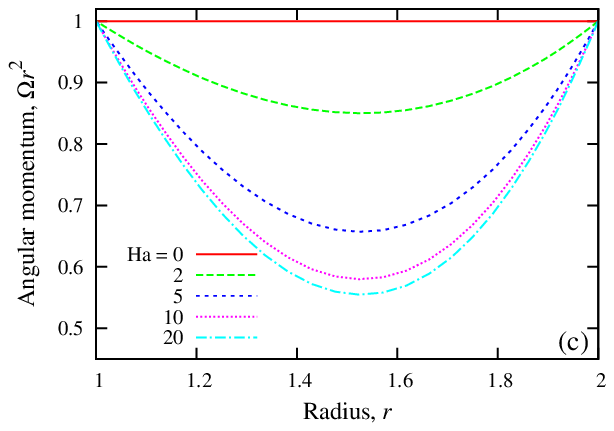} \includegraphics[width=0.45\textwidth]{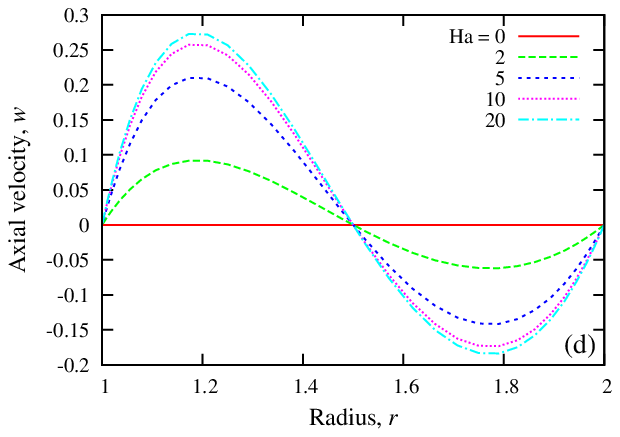}\\
\includegraphics[width=0.45\textwidth]{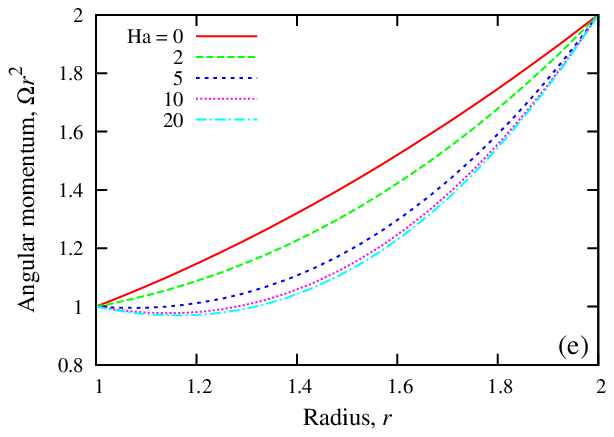} \includegraphics[width=0.45\textwidth]{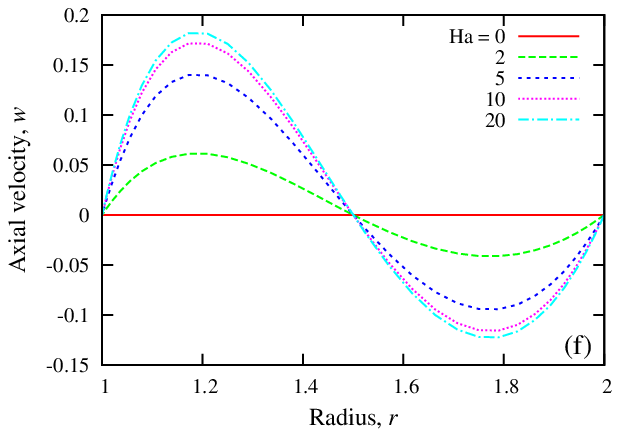}
\par\end{centering}

\caption{\label{fig:mphi-bt6}{[}(a), (c), and (e)] Radial profiles of angular
momentum and {[}(b), (d), and (f)] axial velocity for {[}(a) and (b)]
$\mu=0$ , {[}(c) and (b)] $\mu=0.25,$ {[}(e) and (f)] $\mu=0.5$
and various Hartmann numbers for cylinders with $\lambda=2$ and no
contact resistance $(\bar{\mathcal{R}}=0)$ in helical magnetic field
with $\beta=6.$ }

\end{figure*}

As seen in Fig. \ref{fig:mphi-bt6}, in a helical magnetic field,
the base flow besides the azimuthal component has also an axial one,
which is driven by the interaction of radial current with the azimuthal
component of magnetic field. In the configuration with the endcap
attached to a slower-rotating outer wall, the induced electric current
is flowing radially outward, as discussed above and, thus, the resulting
axial electromagnetic force is directed upward. Because both the current
and azimuthal magnetic field decrease radially outward as $\sim1/r,$
the resulting electromagnetic force is stronger at the inner wall,
where it drives the liquid upward as seen in Figs. \ref{fig:mphi-bt6}(b),
\ref{fig:mphi-bt6}(d), and \ref{fig:mphi-bt6}(f). Since the annular
gap is supposed to be closed at both ends, the constant axial pressure
gradient arising in the response to the electromagnetic force drives
a return flow along the outer cylinder, which compensates for the
upward one along the inner cylinder. This axial flow in the azimuthal
magnetic field, in turn, induces an additional electrostatic potential,
which contributes to that induced by the azimuthal flow in the axial
field as described by Eq. (\ref{eq:dphi0}). The total potential difference
induced by the flow between the inner and outer cylinders balances
that induced by the rotation of bottom in the axial magnetic field,
which is given by Eq. (\ref{eq:dphi1}). The potential balance determines
the magnitude of the induced radial current defined by Eq. (\ref{eq:J0}),
which, in turn, interacts with the magnetic field and disturbs the
flow. Thus, the perturbation of the azimuthal flow is weaker in helical
magnetic field than it is in a purely axial one because a part of
the potential difference is compensated by the axial flow {[}see Figs.
\ref{fig:mphi-bt0}, \ref{fig:mphi-bt6}(a), \ref{fig:mphi-bt6}(c),
and \ref{fig:mphi-bt6}(e)].

\begin{figure*}
\begin{centering}
\includegraphics[width=0.45\textwidth]{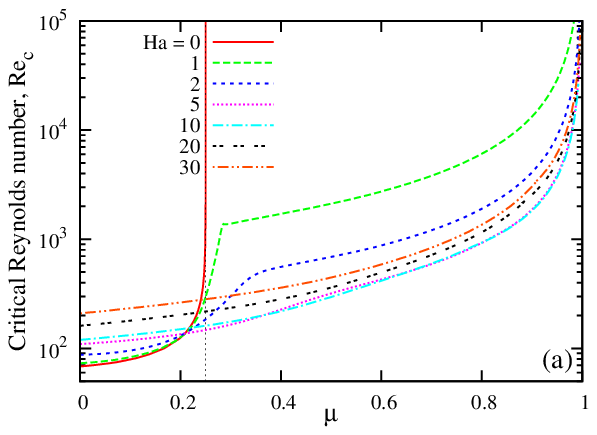} \includegraphics[width=0.45\textwidth]{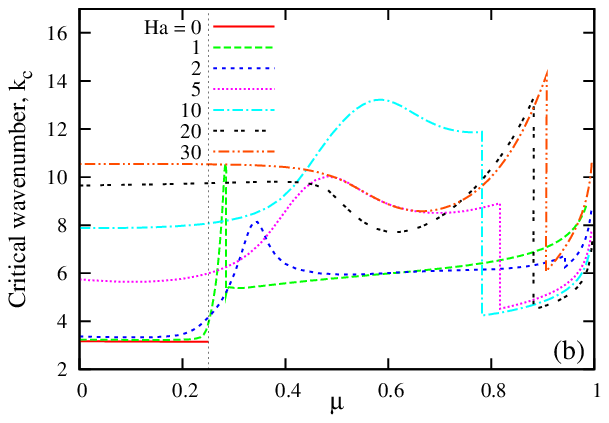}\\
\includegraphics[width=0.45\textwidth]{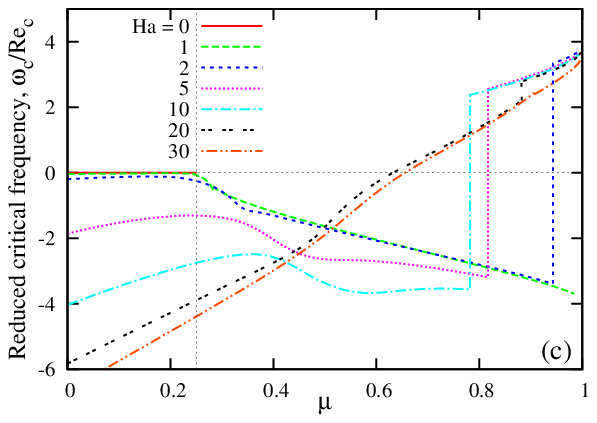}
\par\end{centering}

\caption{\label{fig:Rec-mu-bt6}(a) Critical Reynolds number $Re_{c},$ (b)
wave number $k_{c}$ and (c) the rescaled frequency $\omega_{c}/Re_{c}$
versus $\mu$ at various Hartmann numbers for cylinders with $\lambda=2$
and no contact resistance $(\bar{\mathcal{R}}=0)$ in helical magnetic
field with $\beta=6.$}

\end{figure*}

The instability characteristics in a helical magnetic field plotted
in Fig. \ref{fig:Rec-mu-bt6} differ considerably from those in axial
magnetic field shown in Fig. \ref{fig:Rec-mu-bt0}. In contrast to
the axial magnetic field, now the most unstable mode of instability
is oscillatory, i.e., a traveling wave as for the HMRI. However, it
is important to note that the phase velocity of this wave, which is
determined by the sign of the frequency shown in Fig. \ref{fig:Rec-mu-bt6}(c),
is directed upward oppositely to that of true HMRI. The reversed phase
velocity is due to the longitudinal flow, which is absent for the
ideal HMRI with electrically decoupled cylinders. As seen in Figs.
\ref{fig:mphi-bt6}(a), \ref{fig:mphi-bt6}(c), and \ref{fig:mphi-bt6}(e),
the radial current interacting with the axial component of the magnetic
field causes the angular momentum to decrease radially outward at
the inner cylinder that renders the flow centrifugally unstable. Furthermore,
the Taylor vortices arising at the inner cylinder are advected by
the longitudinal flow upward. The advection in this case obviously
dominates over the direct electromagnetic effect of helical magnetic
field, which would drive the true HMRI wave in the opposite direction.

Moreover, in a helical magnetic field in contrast to purely axial
one, the instability is seen to extend much farther beyond the Rayleigh
line up to the limit of solid-body rotation defined by $\mu=1$ and
even beyond it, which is not considered here. The instability in helical
magnetic field differs significantly from that in purely axial field.
As seen in Fig. \ref{fig:Rec-mu-bt6}, for $\Ha=1,$ shortly after
the Rayleigh line, the most unstable mode switches from the initial
Taylor vortices branch to another one, which is obviously associated
with the axial flow. For larger Hartmann numbers, this transition
proceeds smoothly with the critical wave number developing a maximum
at certain $\mu$ when $\Ha\lesssim10.$ Beyond the Rayleigh line,
the critical Reynolds number first decreases with the Hartmann number
up to $\Ha\approx10$ and then starts to grow for larger $\Ha$ again.
For larger $\mu,$ the most unstable mode jumps to another branch
with a considerably smaller critical wave number and positive frequency,
which corresponds to the opposite direction of the phase velocity
now coinciding with that of the true HMRI. Note that such jumps of
the critical mode are characteristic also for the conventional TD
flow \cite{DiPrima-1959,Hughes-Reid-1964}.

\section{\label{sec:Summ}Summary and conclusions}

We have considered linear stability of a TD-type flow of an electrically
conducting liquid in the annulus between two infinitely long perfectly
conducting and differentially rotating cylinders in the presence of
a generally helical magnetic field. The cylinders were supposed to
be electrically connected through a remote endcap. We showed that
this electrical connection can render the base flow hydrodynamically
unstable. First, the azimuthal base flow in an axial magnetic field
gives rise to a radial emf. If the cylinders are electrically decoupled,
no current can close between them, and, consequently, the emf results
in the radial charge redistribution, which gives rise to the electrostatic
potential whose gradient compensates the original emf. If there is
no current, there is no electromagnetic force and no effect of the
magnetic field on the base flow either. This corresponds to the ideal
TC flow, which is used as a reference for the definition of MRI, where
the magnetic field is expected to destabilize the base flow by affecting
only its disturbances but not the base flow itself. 

This is no longer the case when the cylinders are electrically connected,
and a radial current can close between them. The interaction of radial
current with the axial component of the magnetic field gives rise
to the azimuthal electromagnetic force, which tries eliminate the
velocity difference between the endcap and the liquid above it. Depending
on the strength of magnetic field and the effective contact resistance
between the inner and outer cylinder, this electromagnetic force can
modify the profile of azimuthal base flow so that it becomes centrifugally
unstable. As a result, the magnetic field makes the instability extend
significantly beyond its apparent Rayleigh line so resembling MRI
in the case of an unperturbed TC flow. Furthermore, in a helical magnetic
field, the interaction of radial current with the azimuthal component
of magnetic field gives rise to an axial electromagnetic force, which
drives a longitudinal flow. First, this longitudinal flow going upward
along the inner cylinder, where the azimuthal base flow is destabilized
by the magnetic field, advects Taylor vortices, so giving rise to
a traveling wave as in helical MRI. However, the direction of the
most unstable traveling wave of this centrifugal instability is opposite
to that of the true MRI. Second, for sufficiently large differential
rotation, the longitudinal flow becomes hydrodynamically unstable
itself. For electrically connected cylinders in helical magnetic field,
hydrodynamic instability is possible at any sufficiently large differential
rotation. In this case, there is no pure hydrodynamic stability limit
defined in the terms of the critical ratio of rotation rates of inner
and outer cylinders that would allow one to discriminate between magnetically
modified hydrodynamic instability and the HMRI.

From the experimental point of view, a crucial test for the pseudo--MRI
would be the extension of the Taylor vortex flow beyond the Rayleigh
line in purely axial magnetic field at $\Rm\lesssim1.$ The PROMISE
experiment reports only one such apparently successful test in which,
however, the time-averaged flow and thus stationary Taylor vortices,
if any, are removed \cite{Rued-apjl}. Traveling wave appears as
soon as the azimuthal component of the field is switched on. As to
the helical magnetic field, the experiment \cite{Stefani-NJP} seems
to find the right direction of the phase velocity in agreement with
the ideal HMRI model rather than that of the pseudo--MRI considered
in this paper. But this does not necessarily mean that the real base
flow in the experiment is any closer to the ideal TC one. Note that
the nonaxisymmetric $m=1$ instability mode unexpectedly observed
in the PROMISE experiment is characteristic for certain regimes of
the conventional TD flow \cite{Chen-1993}.

Although the current circulation through the liquid metal has been
eliminated in a modified PROMISE experiment\cite{Stefani-etal2009a}
by insulating the inner cylinder, the base flow still remains strongly
affected by the Ekman pumping due to the endcaps which makes it more
complex than the one used in this study. In the new PROMISE-2 setup
\cite{Stefani-etal2009b,Stefani-etal2009c}, the Ekman pumping has
been significantly reduced by using split rings for the endcap, which
is insulating now, and thus prevents the current circulation through
it. Although the instabilities appear much sharper in the new setup
than in the previous one, the actual hydrodynamic stability limit,
if any, of the base flow and so the nature of the observed instabilities
is still unclear. In particular, as shown by Szklarski and R\"udiger
\cite{Szklarski-Ruediger2007}, the base flow may significantly be
affected by the magnetic field also when the endcaps are insulating
provided that $\Ha\gtrsim10.$

In conclusion, it is not appropriate to use the Rayleigh line of the
ideal TC flow as a criterion to determine the MRI in a significantly
different flow. More elaborate numerical analysis may be necessary
for this purpose.

\begin{acknowledgments}
The author would like to thank Gunter Gerbeth, Frank Stefani and Jonathan
Hagan for constructive comments.
\end{acknowledgments}

\end{document}